# Discovery of dissipative microwave photonic solitons

Tengfei Hao[1,2,3,†], Hao Ding[4,†], Wei Li[1,2,3], Ninghua Zhu[1,2,3], Yitang Dai[4,*], Ming Li[1,2,3,*]

[1]State Key Laboratory on Integrated Optoelectronics, Institute of Semiconductors, Chinese Academy of Sciences, Beijing 100083, China.
[2]School of Electronic, Electrical and Communication Engineering, University of Chinese Academy of Sciences, Beijing 100049, China.
[3]Center of Materials Science and Optoelectronics Engineering, University of Chinese Academy of Sciences, Beijing 100190, China.
[4]State Key Laboratory of Information Photonics and Optical Communications, Beijing University of Posts and Telecommunications, Beijing 100876, China.

†These authors contributed equally to this work.
*Corresponding authors: ytdai@bupt.edu.cn; ml@semi.ac.cn.

**Abstract:**
**Dissipative solitons rely on the double balance between nonlinearity and dispersion as well as gain and loss have attracted a lot of attention in optics [1-9], since it gives rise to ultrashort pulses and broadband frequency combs with good stability and smooth spectral envelopes. Here we observe a novel dissipative solitons in microwave photonics that gives rise to wideband tunable frequency hopping microwave signals with fast frequency switching speed. The dissipative microwave photonic solitons are achieved through the double balance between nonlinear gain saturation and linear filtering as well as gain and loss in a microwave photonic resonant cavity. The generation of dissipative solitons with different pulse width, repletion rate and number of solitons per round-trip time are observed, together with the corresponding wideband tunable frequency hopping microwave signals. This work opens new avenues for signal generation, processing and control based on the principle of solitons in microwave photonics, and has great potential in many applications such as modern radars, electronic warfare systems, and telecommunications.**

**Introduction**

Solitons refers to a physical phenomenon in which a wave packet or pulse that maintains its shape thanks to a certain balance of nonlinear and linear effects. It was first observed in the context of water waves, then also in optics to describe optical fields with a delicate balance between nonlinear and dispersive effects. In particular, dissipative optical solitons generated in nonlinear optical cavities through the double balancing of optical nonlinearity and dispersion as well as gain and loss [1-9] have attracted a lot of attention, since it allows the generation of stable ultrashort pulses and broadband frequency combs with low noise and smooth spectral envelopes. Dissipative optical solitons have also been widely investigated in applications ranging from all-optical buffer [10], photonic and microwave signal synthesizer [11-13], laser metrology and spectroscopy [14-16], to optical communications [17,18]. On the other hand, microwave photonics is an interdisciplinary area that brings together the worlds of optics and radio-frequency (RF) engineering, which has attracted great interest in recent years and has advanced many applications in defence, communication networks, imaging and instrumentations [19-24]. This technology using optical devices and technologies to generate, process, manipulate and distribute RF signals and makes it possible to have functions or performances that are complex or even not achievable by traditional RF systems. Similar with nonlinear optical cavities that could generate light waves in optics, an optoelectronic oscillator (OEO) is a paradigmatic nonlinear microwave photonic cavity that could produce microwave signals [25-29]. Various signals, such as ultra-stable single frequency microwave signals [30-32], linearly chirped microwave waveforms [33], and complex microwave oscillations such as broadband chaotic signals [34-37] have been generated using OEOs. As a hybrid nonlinear microwave photonic resonator, the OEO is benefited from advantages inherent from both optical and microwave domain, such as broad-bandwidth, unprecedented controllability, tunability and precision.

Here we observe the generation of a novel dissipative microwave photonic solitons in OEO that gives rise to wideband tunable frequency hopping microwave signals. The generated dissipative microwave photonic solitons are localized wave packets of microwave field that maintains its shape thanks to the double balance between nonlinear gain saturation and linear filtering, as well as gain and loss in the microwave photonic resonant cavity. The pulse width of the generated dissipative microwave photonic solitons can be easily tailored by changing the filter bandwidth in the microwave photonic resonator, and the repletion rate, number of solitons per round-trip time as well as the frequency spacing of the corresponding frequency hopping microwave signals can also be easily tuned. Our studies show that it is possible to manipulate and tailor microwave photonic systems based on the principle of solitons in a nonlinear microwave photonic cavity, paving the way for a new class of soliton microwave photonic systems for the generation, processing and control of microwave and RF signals.

**Results**

The schematic diagram of the adapted OEO for the observation of dissipative microwave photonic solitons is shown in Fig. 1(a). The OEO is a hybrid microwave photonic nonlinear resonator that consists of an optical part and an electrical part. The optical part is constructed by the use of a laser diode (LD), a phase modulator (PM), an optical fiber, a dual-passband optical notch filter, an erbium-doped fiber amplifier (EDFA) and a photodetector (PD). The electrical part consists of an electrical power divider and an electrical amplifier (EA). Fig. 1(b) shows a simplified version of the OEO loop. As can be seen, the OEO is pumped by an optical signal from the LD. The optical signal and electrical signal in the OEO loop are converted into each other after one cavity round-trip time. Different from conventional dissipative optical solitons that consists of a great deal of cavity modes, only two groups of intercoupled oscillation modes are selected in the OEO cavity to enable the generation of dissipative microwave photonic solitons. An equivalent dual-passband microwave photonics filer (MPF) is incorporated into the OEO cavity to select the desired two groups of intercoupled oscillation modes as well as to provide the linear filtering effect. The equivalent dual-passband MPF is achieved thanks to the joint use of the PM and the dual-passband optical notch filter to achieve phase modulation to intensity modulation (PM-IM) conversion [38,39]. As shown in Fig. 1(c), the shape of passbands of the MPF in the RF domian is converted reversely from the optical notch filter in the optical domain, so the bandwidth of the MPF follows that of the optical notch filter. The center frequencies ($f_1$ and $f_2$) of the two passbands of the MPF are equal to the frequency differences between the optical pump lightwave ($f_c$) and the two notch positions of the optical notch filter, so the center frequencies can be tuned easily by modify either of LD or the optical notch filter. In the adopted OEO configuration, the potential oscillation modes are two groups of intercoupled modes selected by the dual-passband MPF. The delay-differential equation [34] of this OEO can be written as

$$\tau \frac{dx(t)}{dt} + x(t) = G_\text{L} G_\text{NL} x(t-T) \tag{1}$$

where $x(t) = x_1(t) + x_2(t)$ is the oscillation signal in the OEO, $\tau = 1/(\pi \Delta f)$ is related to the bandwidth $\Delta f$ of the MPF, $G_\text{L} = G_A R_\text{PD} I_\text{PD} Z_\text{PD}$ and $G_\text{NL} = 2J_1(\pi/V_\pi \cdot |x(t-T)|)/|x(t-T)|$ are the linear and nonlinear gain saturation factor, respectively. $G_A$ is the gain of the amplifier. $R_\text{PD}$, $I_\text{PD}$ and $Z_\text{PD}$ are the responsivity, input power and impedance of the PD, respectively. $J_n$ is the n-th order Bessel function of the first kind, $V_\pi$ is the half-wave voltage of the modulator and $T$ is the cavity round-trip time of the OEO. According to Eq. (1), we can have the coupled equation of the two groups of oscillation modes

$$\begin{aligned}\tau \frac{dx_1(t)}{dt} + x_1(t) &= \beta \cdot J_0(|x_2(t-T)|) \frac{2J_1(|x_1(t-T)|)}{|x_1(t-T)|} \cdot x_1(t-T) \\ \tau \frac{dx_2(t)}{dt} + x_2(t) &= \beta \cdot J_0(|x_1(t-T)|) \frac{2J_1(|x_2(t-T)|)}{|x_2(t-T)|} \cdot x_2(t-T)\end{aligned} \tag{2}$$

where $\beta = \pi/V_\pi \cdot G_\text{L}$. Under stable oscillation, which means that the gain is balanced with the loss and the oscillation is repeated itself after each round-trip, we can have the equation describing the product of the wave pockets of these two groups of modes

$$\frac{\tau}{2} \frac{dz(t)}{dt} + z(t) = (\beta' + \beta''|z(t-T)|)z(t-T) \tag{3}$$

where $z(t) = x_1(t)x_2^*(t)$ is the product of the wave pockets of these two groups of modes, $\beta' = (3-\beta)/2$ and $\beta'' = 3\beta/8$. The details of the derivation process of Eq. (3) are given in the Supplementary Information. As a delay-differential equation, a straight forward analytical solution of Eq. (3) is still challenging. Nevertheless, we can still analyze the characteristics of $z(t)$ based on this equation. As can be seen, there is a nonlinear gain saturation factor of $\beta''|z(t-T)|$ for $z(t)$. A higher gain can be achieved for those of $z(t)$ with larger amplitude, thus the pulse width of $z(t)$ would be compressed when it travels in the OEO cavity. It should be noted that the nonlinear gain saturation factor for a traditional single mode oscillation in Eq. (1) is decreased with the increase of the signal amplitude, thus the coupling between the two groups of modes are essential to achieve the desired gain saturation effect to compress the pulse width of $z(t)$. At the same time, the pulse width of $z(t)$ is stretched by the linear filtering effect provided by the bandpass MPF since the frequency spectrum is compressed by the MPF. The linear filtering effect is represented by $\tau$ in Eq. (3). As depicted in Fig. 1(d), the nonlinear gain saturation effect is balanced with the linear filtering effect under stable oscillation, so $z(t)$ would maintain its shape. Moreover, the cavity gain provided by the LD and amplifiers must be balanced with the loss in the stable oscillation, since the oscillation would be attenuated by the loss of the MPF at each round-trip. A stable frequency hopping microwave signal with fast frequency switching speed would be generated in this stable operation. Since the delicate double balance between nonlinear gain saturation and linear filtering, as well as gain and loss is similar with that of dissipative optical solitons, we refer this phenomenon as a unique dissipative microwave photonic solitons in the OEO cavity.

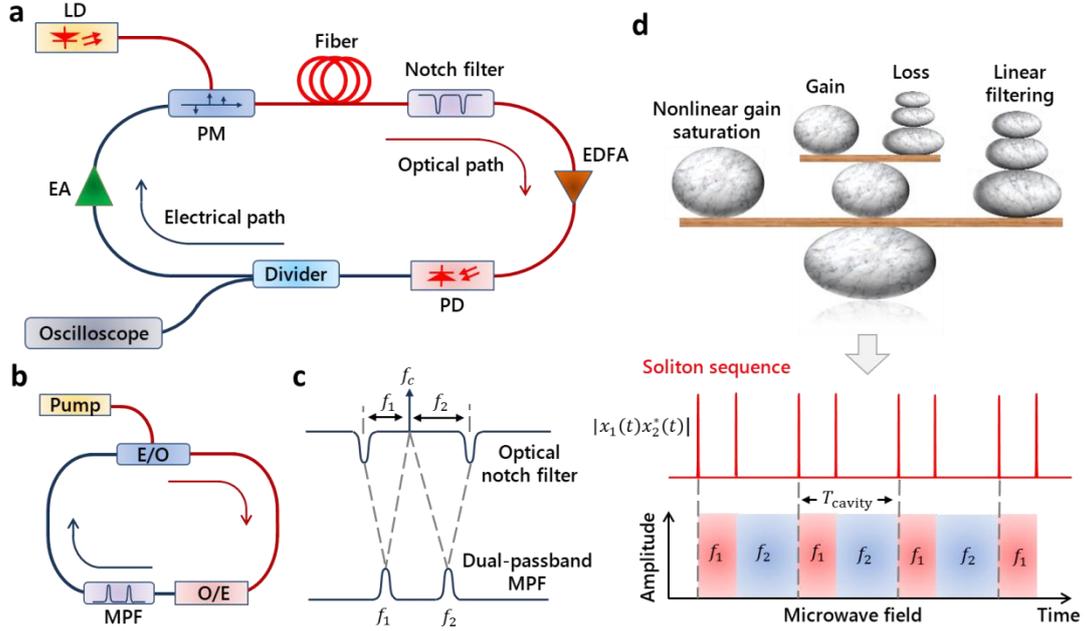

**Fig. 1**. Dissipative microwave photonic solitons in an optoelectronic oscillator (OEO) cavity. (a) Schematic diagram of the adapted OEO for the generation of dissipative microwave photonic solitons. The OEO has a hybrid nonlinear resonant cavity formed with an optical path and an electrical path. (b) Simplified version of the OEO loop. An equivalent dual-passband microwave photonics filter (MPF) is used for the selection of two groups of intercoupled oscillation modes. (c) Principle of the equivalent MPF. The shape of the MPF is converted reversely from the optical notch filter, where $f_c$ is the frequency of the LD, $f_1$ and $f_2$ are the center frequencies of the two passbands of the MPF, respectively. (d) Principle and illustration of the dissipative microwave photonic solitons in the OEO cavity. The dissipative microwave photonic solitons are the product of the wave pockets of the two groups of intercoupled modes, which maintains its shape thanks to the double balance between nonlinear gain saturation and linear filtering, as well as gain and loss. The dissipative microwave photonic solitons gives rise to frequency hopping microwave signals. Whenever a soliton occurs, a frequency jump occurs. LD, laser diode; PM, phase modulator; EDFA, erbium-doped fiber amplifier; PD, photodetector; EA, electrical amplifier. E/O, electrical to optical conversion. O/E, optical to electrical conversion.

We observed the generation of dissipative microwave photonic solitons with different pulse width, repletion rate and number of solitons per round-trip time, together with the corresponding wideband tunable frequency hopping microwave signals based on the schematic diagram of the OEO shown in Fig. 1(a). The gain of the OEO loop is switched on rapidly in order to achieve the required gain saturation effect. The stable dissipative microwave photonic solitons are generated from the OEO loop when the double balance between the nonlinear gain saturation and linear filtering, as well as the gain and loss is achieved. Fig. 2(a) illustrates the generated soliton sequence when a 20-MHz MPF is used. As can be seen, the period of the generated dissipative microwave photonic solitons is equal to the cavity round-trip time of the OEO, so it is repeated itself after each round-trip. There is a pair of solitons during each cavity round-trip. The detail of the soliton is shown in Fig. 2(b). As can be seem, the pulse width of the soliton is about 43 ns, which is related to the bandwidth of the MPF. Figs. 2 (c) and (d) displays the frequency hopping pattern of the two groups of intercoupled oscillation modes, which is extracted from the temporal waveform of the corresponding frequency hopping microwave signal. As can be seen, the amplitude of each mode is almost constant and there is a fast switching on the time boundaries of these two groups of oscillation modes. Moreover, whenever a soliton occurs, a frequency jump occurs. The period and frequency jump times of the frequency hopping signal follows the period and numbers of the solitons, respectively. The frequency switching speed when the wave pocket is changed from 15% to 85 % or vice versa is about 41 ns, which is similar with the pulse width of the solitons.

Fig. 2(e) shows the instantaneous frequency-time diagram of the corresponding frequency hopping microwave signal. The frequency of the frequency hopping modes is determined by the passbands of the MPF in the OEO cavity. Fig. 2(f) shows the measured frequency response of the dual-passband MPF. As can be seen, the gain of the two groups of frequency hopping modes is similar with each other, which ensures the intercoupling between these two groups of modes since they have analogous potential to be oscillated in the OEO cavity. The formation of the stable dissipative microwave photonic solitons in our OEO is a result of a delicate double balance between the nonlinear gain saturation and linear filtering, as well as the gain and loss, which is similar with that of dissipative optical solitons. The solitons in our microwave photonic cavity enables the generation of frequency hopping microwave signals, which reveals a new mode of operation of the microwave photonic resonator.

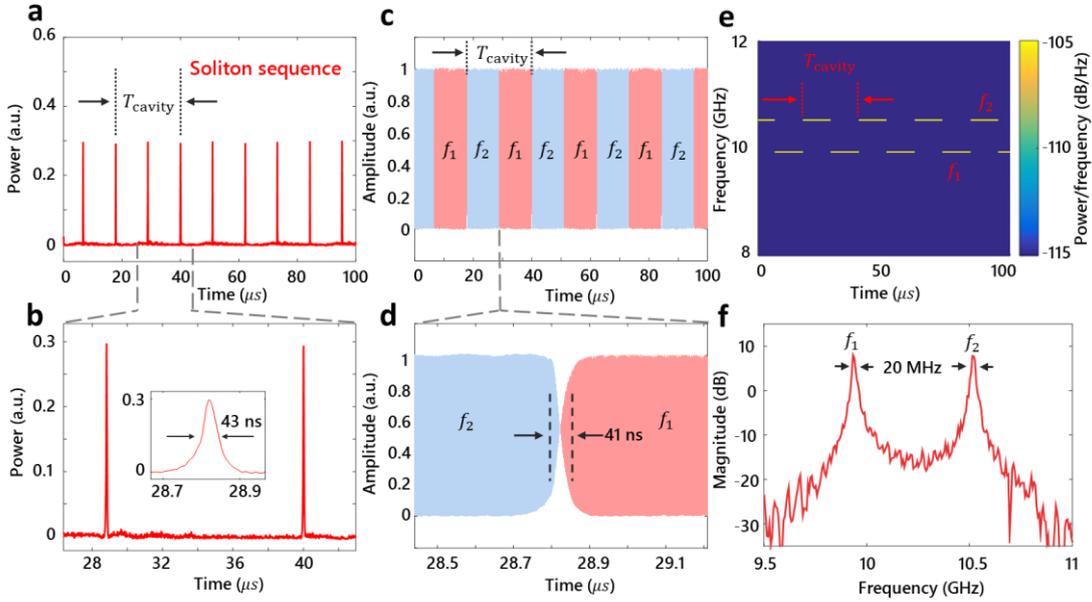

**Fig. 2**. Soliton sequence and corresponding frequency hopping microwave signal of the dissipative microwave photonic solitons when a 20-MHz MPF is used. (a) Generated soliton sequence. (b) Details of the generated solitons. The pulse width of the solitons is about 43 ns. (c) Frequency hopping pattern of the two intercoupled oscillation modes. (d) Details of the frequency hopping pattern. The frequency switching speed of these two modes when the wave pocket is changed from 15% to 85 % or vice versa is about 41 ns, which is similar with the pulse width of the solitons. (e) Calculated instantaneous frequency-time diagram of the frequency hopping microwave signal. (f) Measured frequency response of the 20-MHz MPF.

As Eq. (3) indicates, the pulse width of the microwave photonic solitons is related to the bandwidth of the MPF in the OEO loop. Although a straight forward analytical solution of Eq. (3) is challenging, the pulse width can still be evaluated by altering the experimental parameters and by performing numerical simulation. Figs. 3(a)-(c) shows the generated solitons and corresponding frequency hopping microwave signal when a MPF with a different bandwidth of 50 MHz is used. As can be seen, a shorter soliton pulse width of 18 ns is achieved in this case, which is consistent with our simulation result shown in Fig. 3(b). We also simulated the soliton pulse width when a MPF with different bandwidth is used and the results are shown in Figs. 3(e) and (f). The soliton pulse width is inversely proportional to the filter bandwidth, which can also be deduced from Eq. (3). As a result, the soliton pulse width can be easily tailored by changing the bandwidth of the MPF in the OEO cavity.

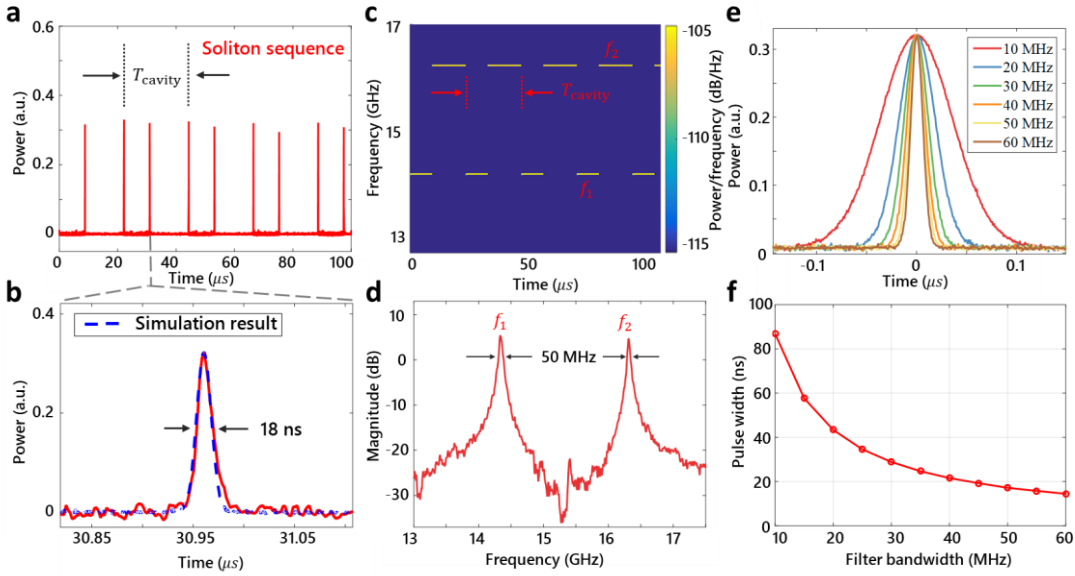

**Fig. 3**. Dissipative microwave photonic solitons when a MPF with different bandwidth is used. (a) Generated soliton sequence when a 50-MHz MPF is used. (b) Details of the generated solitons. Blue dashed line: simulation result. The pulse width of the solitons is about 18 ns, which is shortened compared with the 20-MHz MPF case because of the fall off of the linear filtering effect. (c) Calculated frequency-time diagram of the corresponding frequency hopping microwave signal. (d) Measured frequency response of the 50-MHz MPF. (e) Simulated solitons when a MPF with different bandwidth is used. (e) Relationship between the pulse width of the solitons and the filter bandwidth. The soliton pulse width is inversely proportional to the filter bandwidth.

In addition to the tuning of the soliton pulse width, the repetition rate of the solitons and the frequency spacing of the corresponding frequency hopping microwave signals can also be tuned by simply changing the cavity round-trip time and the center frequencies of the MPF, respectively. In addition, the number of solitons per round-trip may also be varied even when the cavity round-trip time and the passbands of the MPF are both fixed. This can be explained by the fact that although the double balance between the nonlinear gain saturation and linear filtering, as well as the gain and loss enables the generation of dissipative microwave photonic solitons, the number of solitons per round-trip is still not restricted by this operation. The number of solitons per round-trip may be related to the mode competition effect when the oscillation modes get started from noise. Nevertheless, the dissipative microwave photonic solitons are stable once they are established in the OEO cavity. Fig. 4 shows the generated solitons and corresponding frequency hopping microwave signals with different repletion rate, number of solitons per round-trip and frequency spacing of the corresponding frequency hopping microwave signals.

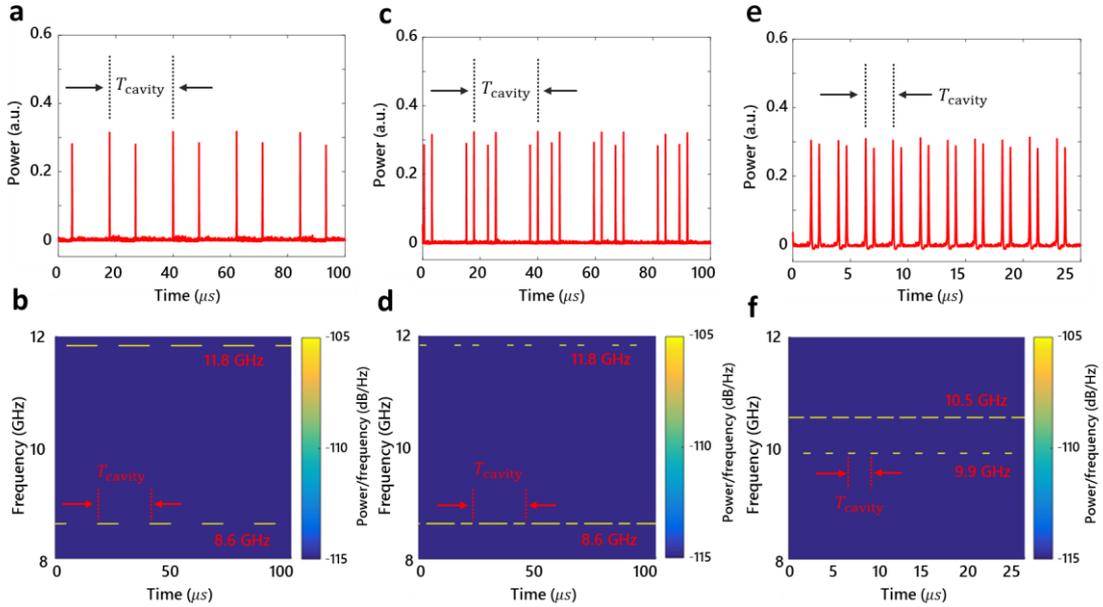

**Fig. 4**. Dissipative microwave photonic solitons and corresponding frequency hopping microwave signals with different repetition rate, number of solitons per round-trip and frequency spacing of the corresponding frequency hopping microwave signals.

**Conclusions**

In conclusion, we discover a novel dissipative microwave photonic solitons that gives rise to wideband tunable frequency hopping microwave signals with a fast frequency switching speed of tens of nanoseconds. The product of the wave pockets of two groups of intercoupled oscillation modes in a microwave photonic resonant cavity is a dissipative soliton since it maintains its shape due to the delicate double balance between nonlinear gain saturation and linear filtering, as well as gain and loss in the microwave photonic cavity. The pulse width of the generated dissipative microwave photonic solitons can be easily tailored by changing the filter bandwidth in the microwave photonic cavity, and the repletion rate, number of solitons per round-trip time as well as the frequency spacing of the corresponding frequency hopping microwave signals can also be easily tuned. This work opens new avenues for signal generation, processing and control based on the principle of solitons in microwave photonics, and has great potential in many applications such as modern radars, electronic warfare systems, and telecommunications.

**Materials and methods**

The adapted OEO for the generation of dissipative microwave photonic solitons was implemented using commercial available optical and optoelectronic devices. The LD is an NKT Koheras Adjustik E15 laser diode with an output power of 18 dBm at 1550 nm. The bandwidth of the PM (EOSPACE) is 40 GHz. The optical fiber is a low dispersion fiber that consists of a single-mode fiber and a dispersion-compensating fiber with opposite dispersion. Two spools of optical fiber with a length of 4.5 km and 450 m were used for repetition rate tuning. An EDFA

(JDS Uniphase) and an EA (SHF 806E) were used to provide sufficient gain to enable solitons generation. The PD is a 15-GHz lightwave converter (HP 11982 A) with a conversion gain of 300 V/W.

The operation of the OEO was evaluated using a digital phosphor oscilloscope (Tektronix DSA73304D) to measure the temporal waveforms and a vector network analyzer (Rohde & Schwarz ZVA40) to measure the frequency response of the dual-passband MPF. The instantaneous frequency-time diagrams were obtained by calculating the short-time Fourier transform of the temporal waveforms.

**Data availability:** The data that support the findings of this study are available from the corresponding author upon reasonable request.

**Conflict of interest:** The authors declare no conflict of interest.

based on phase-modulation to intensity-modulation conversion using a phase-shifted fiber Bragg grating. *IEEE Trans. Microw. Theory Tech.* **60**, 1287–1296 (2012).

# Supplementary Information for

# Discovery of dissipative microwave photonic solitons

Tengfei Hao[1,2,3,†], Hao Ding[4,†], Wei Li[1,2,3], Ninghua Zhu[1,2,3], Yitang Dai[4,*], Ming Li[1,2,3,*]

[1]State Key Laboratory on Integrated Optoelectronics, Institute of Semiconductors, Chinese Academy of Sciences, Beijing 100083, China.
[2]School of Electronic, Electrical and Communication Engineering, University of Chinese Academy of Sciences, Beijing 100049, China.
[3]Center of Materials Science and Optoelectronics Engineering, University of Chinese Academy of Sciences, Beijing 100190, China.
[4]State Key Laboratory of Information Photonics and Optical Communications, Beijing University of Posts and Telecommunications, Beijing 100876, China.

†These authors contributed to this work.
*Corresponding authors: ytdai@bupt.edu.cn; ml@semi.ac.cn.

## 1. Theory of the dissipative microwave photonic solitons

As a delay-line oscillator, the delay-differential equation (DDE) of an OEO can be written as

$$\varphi(t) + \tau \frac{d\varphi(t)}{dt} + \frac{1}{\theta}\int_{t_0}^{t} \varphi(s)\,ds = G\cos^2[\varphi(t-T)] \qquad (S1)$$

where $\varphi(t)$ is the oscillating signal of the OEO, $\tau = 1/(2\pi\Delta f)$, $\Delta f$ is the bandwidth of the loop filter, $\theta = \Delta f/f_0^2$, $f_0$ is the center frequency of the loop filter, $G$ is the gain of the OEO, $T$ is the cavity round-trip time.

When a narrowband loop filter is used, the oscillating signal of the OEO can be considered as a sinusoidal wave

$$\varphi(t) = \frac{1}{2}x(t)e^{i\omega_0 t} + c.c. \qquad (S2)$$

where $A(t)$ and $\omega_0$ are the amplitude and angular frequency of the oscillating signal, respectively. In this case, the DDE of OEO can be further simplified as

$$\tau \frac{dx(t)}{dt} + x(t) = G_L G_{NL} x(t-T) \qquad (S3)$$

where $x(t)$ is the oscillating signal of the OEO, $G_L = G_A R_{PD} I_{PD} Z_{PD}$ and $G_{NL} = 2J_1(\pi/V_\pi \cdot |x(t-T)|)/|x(t-T)|$ are the linear and nonlinear gain saturation factor, respectively. $G_A$ is the gain of the amplifier, $R_{PD}$, $I_{PD}$ and $Z_{PD}$ are the responsivity, input power and impedance of the PD, respectively. $J_n$ is the n-th order Bessel function of the first kind, $V_\pi$ is the half-wave voltage of the modulator and $T$ is the cavity round-trip time of the OEO. It should be noted the value of $\tau$ is doubled compared with the case in Eq. (S1).

In the adapted OEO structure, a dual-passband filter is used to select two groups of intercoupled cavity modes. The total oscillating signal can be expressed as

$$x(t) = x_1(t) + x_2(t) \qquad (S4)$$

where $x_1(t)$ and $x_2(t)$ are the two groups of intercoupled modes selected by the dual-passband filter. According to Eq. (S3) and (S4), the coupled equation of the adapted OEO can be written as

$$\tau\frac{dx_1(t)}{dt} + x_1(t) = \beta \cdot J_0(|x_2(t-T)|)\frac{2J_1(|x_1(t-T)|)}{|x_1(t-T)|} \cdot x_1(t-T)$$
$$\tau\frac{dx_2(t)}{dt} + x_2(t) = \beta \cdot J_0(|x_1(t-T)|)\frac{2J_1(|x_2(t-T)|)}{|x_2(t-T)|} \cdot x_2(t-T)$$
(S5)

where $\beta = \pi/V_\pi \cdot G_L$. According to the properties of Bessel function, we can further simply the coupled equation by ignore the cubic and any other higher order terms

$$\tau\frac{dx_1(t)}{dt} + x_1(t) \approx \beta \cdot \left(1 - \frac{|x_2(t-T)|^2}{4} - \frac{|x_1(t-T)|^2}{8}\right) \cdot x_1(t-T)$$
$$\tau\frac{dx_2(t)}{dt} + x_2(t) \approx \beta \cdot \left(1 - \frac{|x_1(t-T)|^2}{4} - \frac{|x_2(t-T)|^2}{8}\right) \cdot x_2(t-T)$$
(S6)

As can be seen from Eq. (S6), one result of the coupling between the two groups of modes is the introducing of the cross gain saturation effect, where the effective gain for one mode is related to the other mode.

At a steady-state of operation, the oscillating signal would repeat itself after each cavity round-trip, thus we have $x_1(t) \approx x_1(t-T)$ and $x_2(t) \approx x_2(t-T)$. When we consider the product of the wave pockets of the two groups of intercoupled modes, we can have

$$\tau\frac{dx_1(t)}{dt}x_2^*(t) + x_1(t)x_2^*(t) = \beta \cdot \left(1 - \frac{|x_2(t-T)|^2}{4} - \frac{|x_1(t-T)|^2}{8}\right) \cdot x_1(t-T)x_2^*(t-T) \quad (S7)$$

$$\tau x_1(t)\frac{dx_2^*(t)}{dt} + x_1(t)x_2^*(t) = \beta \cdot \left(1 - \frac{|x_1(t-T)|^2}{4} - \frac{|x_2(t-T)|^2}{8}\right) \cdot x_1(t-T) * (t-T) \quad (S8)$$

The sum of Eq. (S7) and (S8) can be expressed as

$$\tau\frac{d[x_1(t)x_2^*(t)]}{dt} + 2x_1(t)x_2^*(t)$$
$$= \beta \cdot \left[2 - \frac{3}{8}(|x_1(t-T)|^2 + |x_2(t-T)|^2)\right] \cdot x_1(t-T)x_2^*(t-T) \quad (S9)$$

At a steady-state of operation, the sum of the amplitude of the two groups of modes is also a constant value $|x_1(t-T)| + |x_2(t-T)| = \sqrt{C}$, so Eq. (S9) can be rewritten as

$$\tau\frac{d[x_1(t)x_2^*(t)]}{dt} + 2x_1(t)x_2^*(t)$$
$$= \beta \cdot \left[2 - \frac{3}{8}C + \frac{3}{4}|x_1(t-T)x_2^*(t-T)|\right] \cdot x_1(t-T)x_2^*(t-T) \quad (S10)$$

As a result, the product of the wave pockets of the two groups of intercoupled modes $z(t) = x_1(t)x_2^*(t)$ should satisfied

$$\frac{\tau}{2}\frac{dz(t)}{dt} + z(t) = \left(1 - \frac{3}{16}C\right)\beta \cdot \left[1 + \frac{3}{8\left(1 - \frac{3}{16}C\right)}|z(t-T)|\right] \cdot z(t-T) \quad (S11)$$

The value of $C$ can be estimated by consider the case when only one mode is activated. In this case, $C$ is the amplitude of the oscillating mode. According to Eq. (S6), we have $C = 8\left(1 - \frac{1}{\beta}\right)$.

As a result, Eq. (S11) can be rewritten as

$$\frac{\tau}{2}\frac{dz(t)}{dt} + z(t) = (\beta' + \beta''|z(t-T)|)z(t-T) \quad (S12)$$

where $\beta' = \frac{3-\beta}{2}$, $\beta'' = \frac{3\beta}{8}$. As the be seen from Eq. (S12), there is a nonlinear gain saturation factor of $\beta''|z(t-T)|$ for $z(t)$. A higher gain can be achieved for those of $z(t)$ with larger amplitude, thus the pulse width of $z(t)$ would be compressed when it travels in the OEO cavity. At the same time, the pulse width of $z(t)$ is stretched by the linear filtering effect provided by the bandpass loop filter since the frequency spectrum is compressed by the filter, which is included into $\tau$ in Eq. (S12). As shown in Fig. S1, the nonlinear gain saturation effect would be balanced with the linear filtering effect under stable oscillation, so $z(t)$ would maintain its shape. Moreover, the cavity gain must also be balanced with the loss in a stable oscillation. Since the delicate double balance between nonlinear gain saturation and linear filtering, as well as gain and loss is similar with that of dissipative optical solitons, we consider this phenomenon as unique dissipative microwave photonic solitons in the microwave photonic cavity.

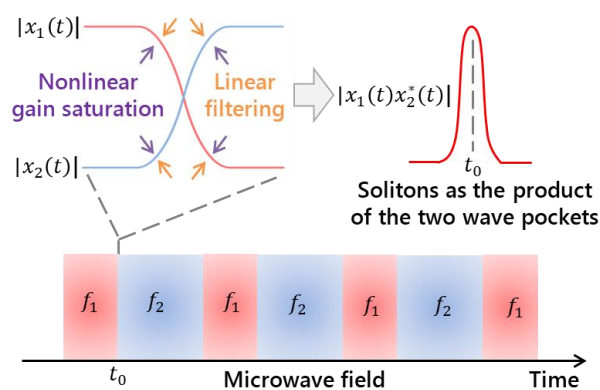

**Fig. S1.** Balance between the nonlinear gain saturation and linear filtering of the dissipative microwave photonic solitons. The pulse width of the product of the wave pockets $z(t) = x_1(t)x_2^*(t)$ is compressed by the nonlinear gain saturation effect and is stretched by the linear filtering effect. The nonlinear gain saturation effect would be balanced with the linear filtering effect under stable oscillation.

2. **Comparison between the dissipative optical solitons and the proposed dissipative microwave photonic solitons**

The key principle of the formation of the dissipative microwave photonic solitons is similar with that of dissipative optical solitons, where they all rely on the double balance between the nonlinear and linear effects as well as gain and loss in a certain resonant cavity. In optical solitons, pulse shaping (compression and broadening) is commonly provided by the nonlinearity and dispersion effects. In microwave photonic solitons, pulse shaping is provided by the nonlinear gain saturation and linear filtering.

At the same time, different from traditional dissipative optical solitons where lots of cavity modes are activated simultaneously, only two groups of intercoupled cavity modes are activated in the dissipative microwave photonic solitons. The product of the wave pockets of the two groups of intercoupled cavity modes is a solition since it maintains its shape due to the double balance mentioned above. As a result, the dissipative microwave photonic solitons gives rise to frequency hopping microwave signals, which may find applications in scenarios such as modern radars, electronic warfare systems, and telecommunications. The temporal waveforms of the corresponding frequency hopping microwave signals measured by a high-speed oscilloscope are shown in Fig. S2.

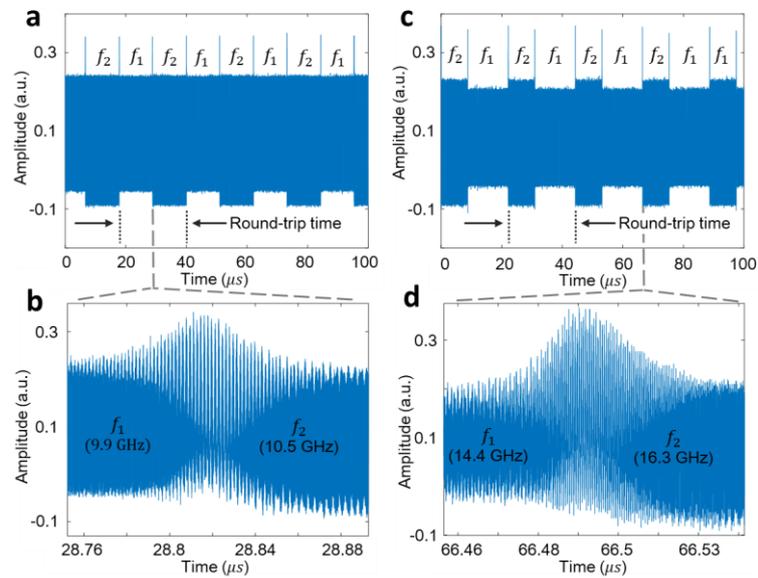

**Fig. S2**. Temporal waveforms of the corresponding frequency hopping microwave signals. (a) The corresponding frequency hopping microwave signal when a 20-MHz MPF is used. (b) Zoom-in view at the boundary of the two groups of intercoupled hopping modes. (c) The corresponding frequency hopping microwave signal when a 50-MHz MPF is used. (d) Zoom-in view at the boundary of the two groups intercoupled hopping modes.